\documentstyle[preprint,aps]{revtex}
\begin{document}
\draft

\title{Violation of Scaling in the Contact Process with Quenched Disorder}
\author{Ronald Dickman$^{\dagger,a}$ and Adriana G. Moreira$^{\ddagger,b}$}
\address{$^{\dagger}$Department of Physics and Astronomy, Lehman College, CUNY,
Bronx, NY 10468-1589 \\ and \\
$^{\ddagger}$Department of Chemistry, SUNY at Stony Brook, Stony Brook, NY 11790-3400}
\date{\today}
\maketitle
\begin{abstract}
We study the two-dimensional contact 
process (CP) with quenched disorder (DCP), and determine the
static critical exponents $\beta $ and $\nu_{\perp}$.
The dynamic behavior 
is incompatible with scaling, as applied to models 
(such as the pure CP) that have a continuous phase transition 
to an absorbing state.  We find that 
the survival probability (starting with all sites occupied), for a 
finite-size system at critical, decays according to a power law, 
as does the {\em off-critical}  density 
autocorrelation function.  Thus the critical exponent
$\nu_{||}$, which governs the relaxation time, is undefined,
since the characteristic relaxation time is itself
undefined.  The logarithmic time-dependence found in recent 
simulations of the critical DCP [Moreira and Dickman, Phys. Rev. E{\bf 54}, 
R3090 (1996)] is further evidence of violation of scaling. 
A simple argument based on percolation cluster statistics
yields a similar logarithmic evolution.
\vspace {0.3truecm}

\end{abstract}
\vspace{1.0truecm}

\noindent $^a${\small e-mail address: dickman@lcvax.lehman.cuny.edu } \\
$^b${\small e-mail address: dri@fisica.ufmg.br }
 
\newpage
\section{Introduction} 

Phase transitions between an absorbing state  
(one admitting no further evolution), 
and an active regime occur in models of  
autocatalytic chemical reactions, epidemics, 
and transport in disordered media \cite{RDPRV}.  
Paradigms of this sort of transition are the contact process (CP)\cite{TEHARRIS}
and its simultaneous-update counterpart, directed percolation (DP) \cite{KINZEL}.
Since many-particle systems often incorporate 
frozen-in randomness, it is natural to investigate the effect of quenched
disorder on an absorbing-state transition.  Thus, some years ago,
Noest observed that the critical behavior of  
disordered directed percolation is quite different from that of pure DP \cite{NOEST}.   
We recently studied the CP with quenched disorder in the form of random site
dilution, and found logarithmic time-dependence at the 
critical point \cite{DCP1}.  For example, $P(t)$, the probability of
survival, starting from a single active site, follows $P \sim 1/(\ln t)^a$ for large $t$.
Such a form is incompatible with the scaling hypothesis 
that applies quite generally to absorbing-state transitions \cite{GRASS}.
This violation of scaling is consistent with Janssen's recent
field-theoretic analysis, which shows that the resulting renormalization
group equations have only runaway solutions \cite{JANSSEN97}
Here we present further results bearing on the violation of scaling.  In particular, 
we find that the exponent $\nu_{||}$ does not exist for the diluted contact
process (DCP).  We also propose an explanation for logarithmic
behavior at the critical point.

In the CP on the square lattice,
each site is either vacant or occupied by a particle.
Particles are created at vacant sites at rate $\lambda n /4$, where $n$ is
the number of occupied nearest-neighbors, and
are annihilated at unit rate, independent of the surrounding
configuration. The order parameter is the particle
density $\rho$; it vanishes in the vacuum state, which is absorbing.
As $\lambda$ is increased beyond $\lambda_c = 1.6488(1)$,
there is a continuous phase transition from the vacuum 
to an active steady state; for $\Delta \equiv \lambda - \lambda_c > 0$, 
the stationary density
$\overline{\rho} \sim \Delta^{\beta}$.  In the vicinity of
the critical point the characteristic relaxation time 
$\tau \sim |\Delta |^{-\nu_{||} } $, and the
correlation length diverges as $\xi \sim |\Delta |^{-\nu_{\perp} } $.
In the DCP, a fraction $x$ of the sites are diluted at random, and the 
birth-and-death process defining the CP is restricted to non-diluted sites.
(Further details on the DCP may be found in Ref. \cite{DCP1}.)

To provide the necessary background we summarize the scaling behavior of
the CP and allied models \cite{GRASS}.  Consider first 
the evolution from an initial configuration with just a single particle at the origin.
The conditional probability of finding a particle at $r$,
given that at time zero there was a particle at the origin, and that
all other sites were vacant, obeys

\begin{equation}
\label{densc}
\rho(r,t;0,0) \simeq t^{\eta - dz/2} F(r^{2}/t^{z}, \Delta t^{1/\nu_{||}}).
\end{equation}

\noindent Similarly the survival probability is expected to follow
\begin{equation}
\label{survsc}
P(t) \simeq t^{-\delta} \Phi(\Delta t^{1/\nu_{||}}).
\end{equation}

\noindent ($F$ and $\Phi$ are scaling functions.)
At critical ($\Delta = 0$), equation (\ref{survsc}) implies
$P(t) \sim t^{-\delta}$,
while (\ref{densc}), when integrated over space, yields a mean
population $n(t) \sim t^{\eta}$.
If we take the second moment 
(in space) of $\rho (r,t;0,0) $ with $\Delta = 0$, we obtain

\begin{equation}
R^2(t) \sim \frac  {t^{\eta - dz/2}}{n(t)} \int x^2 f(x^2/t^z) \; d^{d}x \; \propto \; t^z \;.
\end{equation}
For $\Delta > 0$, the survival probability attains a finite asymptotic value:
$\lim_{t \rightarrow \infty} P(t) \equiv P_{\infty} \sim \Delta^{\beta'}$ 
\cite{betanote,MENDES}.
Several scaling relations can be derived, in particular,
\begin{equation}
z = 2 \nu_{\perp}/\nu_{||},
\end{equation}
and
\begin{equation}
\delta = \beta/\nu_{||} \; ,
\end{equation}
as well as the hyperscaling relation
\begin{equation}
4 \delta + 2 \eta = dz.
\label{hyper}
\end{equation}

Eqs. (1) - (3) describe spreading from a single seed.  Consider,
on the other hand, a system at critical with all $L^d$ sites initially occupied, 
and let $P_m (t) $ be the survival probability starting from this
maximally occupied state.  At short times $P_m =1$ and the density
is governed by the power law $\rho (t) \sim t^{-\delta}$.
Following this initial phase, $P_m$ decays exponentially, with a characteristic time 
$\tau \sim L^{\nu_{||}/\nu_{\perp} } $, and the density
in the {\em surviving} sample attains a quasi-stationary value 
$\rho_s \sim L^{-\beta/\nu_{\perp} } $. 
The scaling results summarized above have been amply confirmed for the CP
and other models with a unique absorbing configuration, and have been extended
to models possessing multiple absorbing configurations \cite{MENDES,MUNOZ}.

In our recent study of the DCP, we found that $P(t)$, $n(t)$ 
and $R^2 (t)$ display logarithmic time dependence at critical, which is 
incompatible with the scaling forms,
Eqs. (1) and (2), describing the pure model.  In this work we report further results
on static and dynamic critical properties, in an effort to determine the extent of
the scaling violation, and to understand its origin.  In Sec. II we analyze the stationary
density (in the supercritical regime), and the quasi-stationary density (at critical), to
obtain estimates of $\beta$ and $\nu_{\perp}$, and also examine the stationary
density-density correlation function, which decays algebraically.  Sec. III concerns 
the survival probability, $P_m$,
starting from a maximally occupied state; it also decays algebraically,
roughly as predicted by a simple probabilistic picture.  Another simple argument is
presented in Sec. IV, for the logarithmic decay of the survival probability.  We conclude,
in Sec. V, with a discussion of our main results, and of the reason for violation of
dynamic scaling in the DCP.

\section{Static Behavior}

\subsection{Stationary Density}

We determined the stationary density $\overline{\rho}$ 
for dilutions $x = 0.05$, 0.1, 0.2, 0.3 and 0.35,
by the following simulation procedure.  After generating a disorder configuration on a
lattice of $L \times L$ sites, and initially occupying all non-diluted sites,  we permit
the system to relax for a time $t_R$, and then accumulate data on the density 
for a period of duration $t_S$.  This process is repeated $N_T$ times (with a new
disorder configuration for each trial), and the density $\rho (t)$,
computed over the $N_S$ trials that survive up to time $t_R + t_S$, is examined
to verify that sufficient time has been allowed for relaxation.  (If not, $t_R$ is
increased accordingly.)  We then take the mean density for each trial during
the observation time $t_R \leq t \leq t_S$, and compute the mean and standard 
deviation over the sample of $N_S$ independent trials.  We used lattice sizes
$L = 100$, 200, 400, and 800, increasing $L$ as we approached the critical point,
and checking (for $L \leq 400$) that estimates for the stationary density agreed
for two lattice sizes.  The sample size $N_S$ ranged from 50 to 100. 
We employed $t_R = t_S = 5 \times 10^4 $ - $2.5 \times 10^5 $,
the larger values reflecting slower relaxation near $\lambda_c (x)$.
(We use the critical point estimates $\lambda_c (x)$ obtained via
time-dependent simulations in Ref. \cite{DCP1}.)

Our results for the stationary density, shown in Fig. 1, 
reveal a crossover
between the DP value of $\beta \simeq 0.58$ at small dilutions 
($x = 0.05 $ and 0.1), and a larger exponent
as one approaches $\lambda_c$.
The data for $x \geq 0.1$ yield exponents in the range 0.89 - 0.98,
suggesting that $\beta = 0.93(5)$.  (Figures in parentheses denote
statistical uncertainties --- one standard deviation.)
Similarly, an analysis of the stationary mean-square
density supports $\overline{\rho^2} \sim \Delta^{2\beta}$ with $2 \beta = 1.84(4)$.
Thus the order parameter exponent
agrees, to within uncertainty, with our earlier result $\beta' = 0.99(3)$ for
the exponent governing the ultimate survival probability \cite{DCP1}.  
(For the pure CP, $\beta' = \beta \simeq 0.58$.)  While our estimate is
not far from Noest's result, $\beta = 1.10(5)$, we regard it as excluding that value;
we do not feel, however, that our data rule out $\beta = 1$.  

\subsection{Quasi-Stationary Density}

Rather than studying spatial correlations directly, we determine $\nu_{\perp}$
by analyzing the quasi-stationary density $\rho_s$ at critical as a function of  
the system size; this yields $\beta/\nu_{\perp}$, as noted in Sec. I.
We studied lattice sizes varying
from $L=8$ to $L=128$, averaging over
$2\times 10^3$ to $10^5$ independent runs, of duration
$t_s=10^3$ to $t_s=10^5$.  (The larger sample sizes and longer run times apply
to the larger $L$ values.)
We show in Fig. 2 a log-log plot of $\rho_s (\lambda_c, L)$ {\it versus} $L$
for dilutions ranging from 0.02 to 0.35, along with the slopes of linear
least-squares fits to the data for $L \geq 16$. (The uncertainty in the slope
ranges from 1 to 3\%.)
For $x < 0.1$, the scaling of $\rho_s$ is similar to that found in the pure CP, 
for which $\beta/\nu_{\perp} = 0.80(3)$. But for larger dilutions 
we see a steady increase in the slope; we estimate 
$\beta/\nu_{\perp} =0.93(3)$ for the DCP.  Combining this with our 
estimate $\beta = 0.93(5)$, 
we obtain $\nu_{\perp} = 1.00(9)$ for the DCP.  This is significantly 
lower than, but still
consistent with, Noest's earlier estimate of 1.17(10) \cite{NOEST}.
On the other hand, a theorem of Chayes et al.
requires $d \nu_{\perp} \geq 2$ for models with quenched disorder, or
$\nu_{\perp} \geq 1$ here \cite{CHAYES}; the upper range of our
estimate is consistent with this result.  (Note that if we use our earlier estimate,
$\beta' = 0.99(3)$ in place of our $\beta$ value, we obtain $\nu_{\perp} = 1.06(7)$.)

\subsection{Stationary Density-Density Correlation Function}

The stationary density-density correlation function is defined by

\begin{equation}
C(t) = \langle \sigma_i (t_0+t) \sigma_i (t_0) \rangle  \;,
\label{defc}
\end{equation}
where $\sigma_i (t) = 1$ (0) if site $i$ is occupied (vacant) at time $t$, and the average is
over realizations of the process {\em and} over disorder.  It is understood that $t_0$ is
sufficiently large that the r.h.s. is independent of $t_0$.  For the undiluted CP, the r.h.s.
is then independent of $i$ as well, and 
$\Delta C \equiv C(t) - \overline{\rho}^2 \sim \exp(-t/\tau) $ for large $t$.  
$\tau$ is a characteristic
relaxation time diverging as $\tau \sim \Delta^{-\nu_{||}} $ near the critical point.

We studied the density-density correlation function in the DCP at $x=0.1$ and 0.3, evaluating
the r.h.s. of Eq. (\ref{defc}) for a {\em single} site $i$ (the first non-diluted site to be generated),
on $L \times L $ lattices with periodic boundaries, using 500 --- 2000 independent
realizations of the disorder.  Fig. 3 (inset) shows a typical evolution, with $C(t)$ slowly 
approaching an asymptote, $C_{\infty}$.  The main graph shows that the excess, 
$\Delta C \equiv C(t) - C_{\infty} \sim t^{-b}$, so it cannot be characterized
by a relaxation time.  The exponent $b$ varies from about 0.7 - 0.8 well above $\lambda_c$,
to about 0.5 near $\lambda_c$.  Note that for $b \leq 1$, even the alternative expression
\begin{equation}
\tau \equiv \int_0 ^{\infty} dt \Delta C(t) \;,
\end{equation}
similar to that employed by Noest, is undefined.
We also observe a power-law approach (with an exponent
of about 1/2) of the critical quasi-stationary density $\rho_s$, discussed in the
preceding subsection, to its asymptotic value.

\section{Dynamic Behavior}

Consider the pure CP on a lattice of $L^d$ sites, starting with all sites occupied.  
Since we are dealing with a finite system, there is a well-defined lifetime $\tau(\Delta, L)$
and at long times the survival probability $P_m (t) \propto \exp[-t/\tau(\Delta,L)]$
for large $t$.  
Just at the critical point the lifetime has a power-law dependence on $L$: 
$\tau(0,L) \sim L^{\nu_{||}/\nu_{\perp}}$.  
For $\Delta > 0$, vacating $(L/\xi)^d$ independent regions simultaneously is an 
exponentially rare event, and we expect the scaling form
\begin{equation}
\tau(\Delta ,L) \sim L^{\nu_{||}/\nu_{\perp}} \exp[c(L \Delta^{\nu_{\perp}})^d] , 
\label{tausc}
\end{equation}
(c is a constant),
as is confirmed by the data shown in Fig. 4.  (The data also appear to scale for
$\Delta < 0$, but with a different scaling function.)

We studied the relaxation
from a maximally occupied state (all non-diluted sites occupied) in the 
critical DCP.   Figs. 5 and 6 show $P_m (t)$
for $x=0.1$ and $0.3$, respectively, for $L= 32$, 64, and 128.  (We studied samples of
5000, 2000, and 1000 trials for $L= 32$, 64, and 128, resp.)  From 
the figures it appears that following the initial stage, the survival 
probability crosses over to a nonuniversal
power law, with an exponent that decreases with $L$.  (For $x=0.1$ we find 
$P_m \sim t^{-a}$ with $a = 1.8$, 1.1, and 0.86 for $L= 32$, 64, and 128, resp.; 
for $x=0.3$ the corresponding powers are 0.71, 0.55, and 0.37.)  Since the 
asymptotic decay of $P_m$ at critical is nonexponential,
there is no characteristic lifetime for the process. The initial stage, 
during which $P_m = 1$, is characterized by a correlation length $\xi (t) < L$.
During this phase we find $n_m \sim t^{-\delta} $ with $\delta \simeq 0.47$, as for the
pure model.  Once $P_m$ starts to decay, $n_m$ crosses over to a different, nonuniversal
power law, as seen in Fig. 7.  

We note in passing that while the $P_m $ data are certainly inconsistent with
exponential decay, a slight (downward) curvature suggests a faster 
than power-law decay.  In fact, somewhat more linear plots are obtained 
using the form $ \ln P_m \propto (\ln t)^{-a} $, with $a$ in the range 1.5 --- 3 depending 
on the data set.  The data are too noisy to permit a definite conclusion regarding the
form of $P_m$, but it is interesting that a simple argument yields the modified power law.
We begin by supposing that because of fluctuations in the disorder, the probability
of a lattice of $L^d$ sites having an effective distance from criticality of $\Delta$ is
$P(\Delta) \sim \exp(- b L^d \Delta^2)$.   (Here we ignore the spatial inhomogeneity of the
disorder, and invoke the central limit theorem.)  If we then
assume that such a lattice is characterized by a lifetime
$\tau (\Delta, L)$ scaling as in the pure model, Eq. (\ref{tausc}), 
the survival probability is given by
\begin{equation}
P_m(t) \sim \int d \Delta \exp \left( - b L^d \Delta^2 - \frac {t}{\tau (\Delta , L)} \right).
\label{psc}
\end{equation}
Maximizing the argument of the exponential to extract the leading behavior at large $t$,
we obtain $\ln P  \sim (\ln t)^{2/d\nu_{\perp}}$.  (The exponent $2/d \nu_{\perp} \simeq 1.37$ 
for $d = 2$.)  

This argument, which treats fluctuations in the disorder as if they
were spatially homogeneous, and uses pure-model scaling, is clearly 
inadequate to deal with the true subtlety of the DCP.  That it
may yet contain some germ of truth is suggested by our finding that
when we do not average over disorder, the decay of $P_m$ is {\it exponential},
as in the pure CP, but with a lifetime particular to the disorder set 
generated.  (One is naturally interested
in knowing the distribution of the relaxation time.  This poses a formidable 
numerical task that we hope to address in future work.)
In summary, the relaxation of the DCP from a maximally occupied state
is similar to that of the pure model during the initial stage, in which correlations have
yet to grow to the size of the system.  But afterward the evolution follows nonuniversal
power laws (or modified power laws) and we cannot define a characteristic lifetime.

\section{Critical Dynamics}

In this section we propose a simple explanation for how logarithmic time-dependence 
arises in the critical DCP.  Consider the survival probability $P(t)$ starting from
a single occupied site or seed located at the origin, ${\cal O}$.  Clearly,
the trials that contribute to $P(t)$ at large $t$ are those in which the seed happens to
fall in a large, favorable region.  To make the notion of a ``favorable region" somewhat
more precise, imagine taking the disorder configuration on a cube of $L^d$ sites,
filling space with periodic copies, and running the contact process with 
$\lambda = \lambda_c (x) $ on this lattice.  For some disorder configurations --- the
favorable ones --- the process will in fact be {\em supercritical}, because the fraction
of diluted sites is $ < x$, or because of a particularly advantageous arrangement of
the diluted sites.  Such regions are characterized by an ``effective distance from
criticality" $\Delta_{eff} > 0$.
Any disorder configuration contains both favorable and unfavorable
regions.  If the seed lies in a favorable region, we can define its ``domain" as the 
maximal connected favorable region containing ${\cal O}$.  (The domain is
surrounded by unfavorable regions, which impede the spread of the process.)
For simplicity, we suppose that on a domain of $V$ sites, 
having some $\Delta_{eff} > 0$, the DCP has a lifetime 
$\tau \sim \exp [c V \Delta_{eff}^{d \nu_{\perp}}] $, as it would on a compact
region, in the pure model.  From the central limit theorem, the  
typical value of $\Delta_{eff}$ on a domain of $V$ sites $\sim V^{-1/2}$, yielding a
lifetime $\tau \sim \exp[c V^{1- d\nu_{\perp}/2}] $.  At time $t$, only trials 
whose seeds happen to fall in a domain
with $\tau \geq t$, or $V \geq c'(\ln t)^{2/(2-d \nu_{\perp})} $ survive. 

It remains to estimate the probability $p(V)$ that ${\cal O}$ belongs to a domain of $V$
sites.  To do this, note that a domain is a kind of percolation cluster.  The precise definition of
the sites in this percolation problem is unclear (we might imagine averaging over small
blocks of sites in the original lattice), as is the connectivity rule (next-nearest neighbor
blocks, for example, might effectively be connected).  But we should expect the associated
percolation model to be isotropic and of finite range.  Moreover, domain percolation must
be {\em critical} at $\lambda_c (x)$.  If it were supercritical, the contact process would be
able to spread into an unbounded domain, and so would itself be supercritical.  Similarly,
if domains were subcritical, their size distribution would decay exponentially, and the CP
would be subcritical.  At critical, the domain size is power-law distributed: 
$p(V) \sim V^{-(\tau_p - 1)} $ for large $V$, with $\tau_p$ the usual percolation cluster-size
exponent ($\tau_p = 187/91$ in two dimensions).  Combining this result with the lifetime estimate,
we have
\begin{equation}
P(t) \sim \int_{c'(\ln t)^{2/(2-d \nu_{\perp})} } \frac {dV}{V^{\tau_p -1}} 
\sim (\ln t)^{-2(\tau_p -2)/(2-d \nu_{\perp})} .
\label{logsca}
\end{equation}
As in the argument (Sec. III) for the survival probability 
starting from a maximally
occupied state, the effect of inhomogeneity in the disorder is greatly oversimplified.
Inserting the known values of $\nu_{\perp}$ and of $\tau_p$, we obtain 
$P(t) \sim (\ln t)^{-0.2} $, whereas the exponent we observed in simulations \cite{DCP1}
is much larger, and {\em nonuniversal,} ranging from about 2.7 at $x=0.35$ to
4.6 at $x=0.1$.
Thus we offer the above argument without any claim of quantitative
validity, but rather to show how a simple treatment of disorder leads naturally
to logarithmic time-dependence, and in the hope that it may form
the basis for a more convincing approach. 

The probabilistic arguments suggest that it may be possible to
understand how anomalous dynamics arises from an average over disorder.
Here it is important to recall Noest's analysis
of contributions from exponentially rare, favorable disorder configurations 
to the survival probability $P(t)$.
By deriving upper and lower bounds on the survival probability, he was able to prove 
power-law decay of $P(t)$ in a Griffiths phase for 
$\lambda > \lambda_c (0) $, in the nonpercolating regime \cite{NOEST88}.  Arguments
of a somewhat similar nature were advanced by Bray in his discussion of the relaxation
of diluted spin models \cite{BRAY}.

\section{Discussion}

We have found that some aspects of the diluted contact process
exhibit the same sort of critical behavior --- albeit with different exponents --- as
seen in the pure model.  Other features --- spreading from a seed,
the density-density autocorrelation function, $C(t)$, and the survival
probability $P_m(t)$, starting from a maximally occupied 
state --- do not follow the usual scaling, and are nonuniversal.   
The anomalous properties are all connected
with dynamics:  normally, $C(t)$ and $P_m (t) $ decay exponentially, with the
diverging lifetime serving to define the exponent $\nu_{||}$ 
through $\tau \sim \Delta^{-\nu_{||}}$.
Here no such definition is possible.  Consistent with this, the spreading exponents
$\delta $, $z$, and $\eta$, which are formally zero, are connected to $\nu_{||}$ via
the scaling relations Eqs. (4) --- (6).  Since $\beta$ and $\nu_{\perp}$ are in fact finite,
Eqs. (4) and (5) suggest that $\nu_{||}$ is infinite.
Given Janssen's recent results \cite{JANSSEN97}, it is of interest to 
know whether our finding of power-law
static behavior, but anomalous time-dependence, is compatible with a field-theoretical
analysis .

Some insight into the violation of dynamic scaling may be gained by returning to
Eq. (\ref{tausc}): the exponential dependence of the lifetime (in the pure CP) upon
$L$ and $\Delta$ suggests an extreme sensitivity of dynamic behavior to
disorder.  We might expect dynamics to be dominated by the extremes of local
fluctuations in the disorder.  The result is that a property such as the relaxation time
$\tau$ for the density-density autocorrelation function is {\em non-self-averaging},
i.e., it does not converge to a limiting value even as 
$L \rightarrow \infty$ \cite{WISEMAN,AHARONI}.
Another manifestation of non-self-averaging is the dependence of $P(t)$ and $n(t)$,
even at long times, on the location of the seed at time zero.  These features are
dominated by local fluctuations, rather than by the properties of a ``typical" disorder
configuration.  This in turn suggests that further insight may be gained by studying
critical behavior for {\em fixed} disorder configurations, in order to determine the
statistical distributions of various system properties, and of the domains defined in
Sec. IV.  In this way the primitive probabilistic arguments for $P_m$ and $P(t)$ 
could be honed into a quantitative description of how anomalous behavior 
emerges in the average over disorder.
\vspace{1em}

\noindent{\bf Acknowledgment}
\vspace{1em}

A.G.M. is grateful for support by CAPES (Brazil).

\newpage

\noindent{\bf Figure Captions}
\vspace{1em}

\noindent FIG. 1.  Stationary density, $\rho$, versus $\Delta \equiv \lambda - \lambda_c$ in the diluted contact process. $\times$: $x=0.05$; filled squares: $x=0.10$; $\circ$: $x=0.20$;
$\Box$: $x=0.30$; $\bullet$: $x=0.35$.  Figures denote the slopes of the various straight
lines.
\vspace{1em}

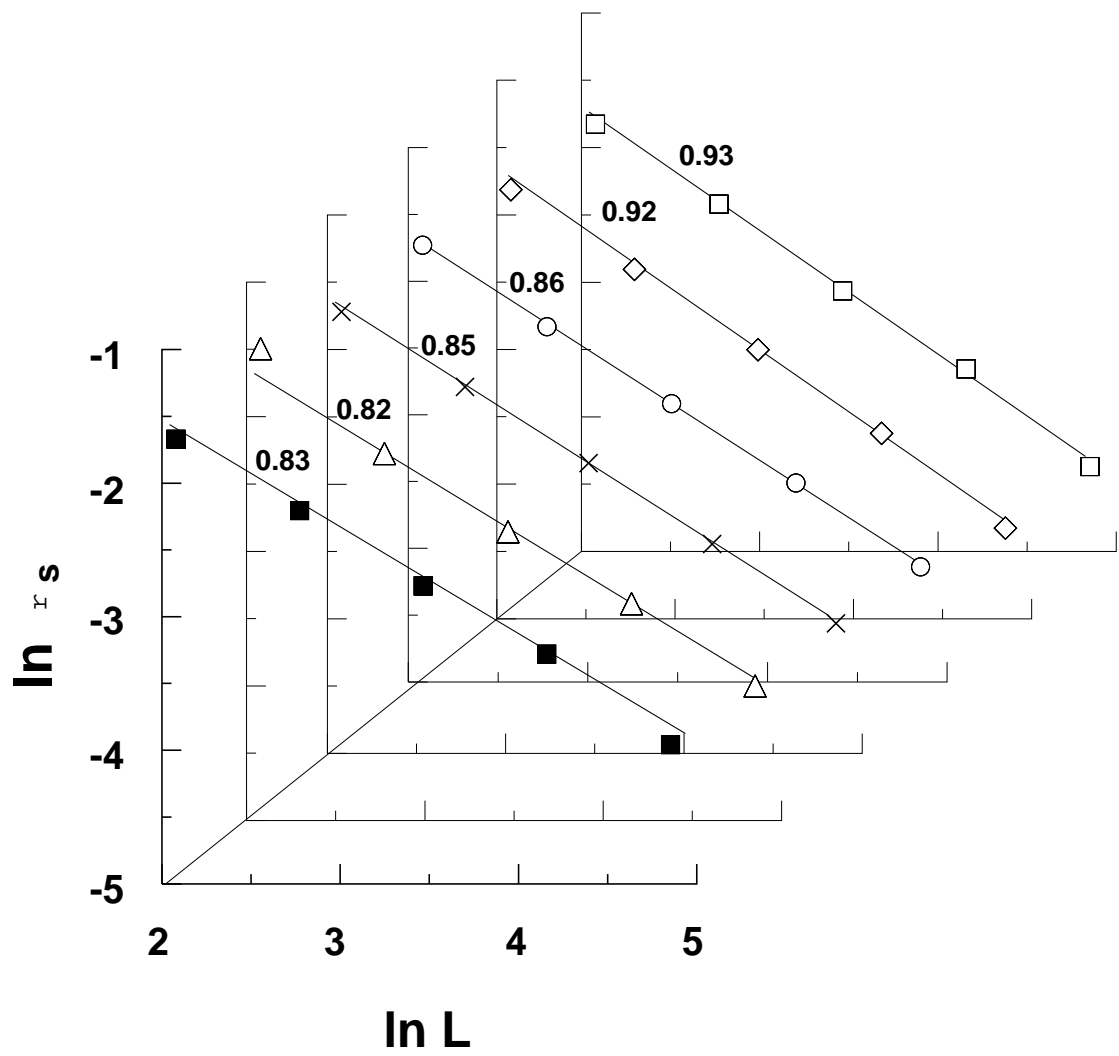
\noindent FIG. 2. Quasi-stationary density $\rho_s (\lambda_c,L)$ {\it versus} $L$,
for dilutions $x= 0.02$ (filled squares); 0.05 ($\triangle$); 0.10 ($\times$); 0.20 ($\circ$); 
0.30 ($\diamond$) and 0.35 ($\Box$).  The lines are linear fits to the last four
data points in each set; figures indicate the slope.
\vspace{1em}

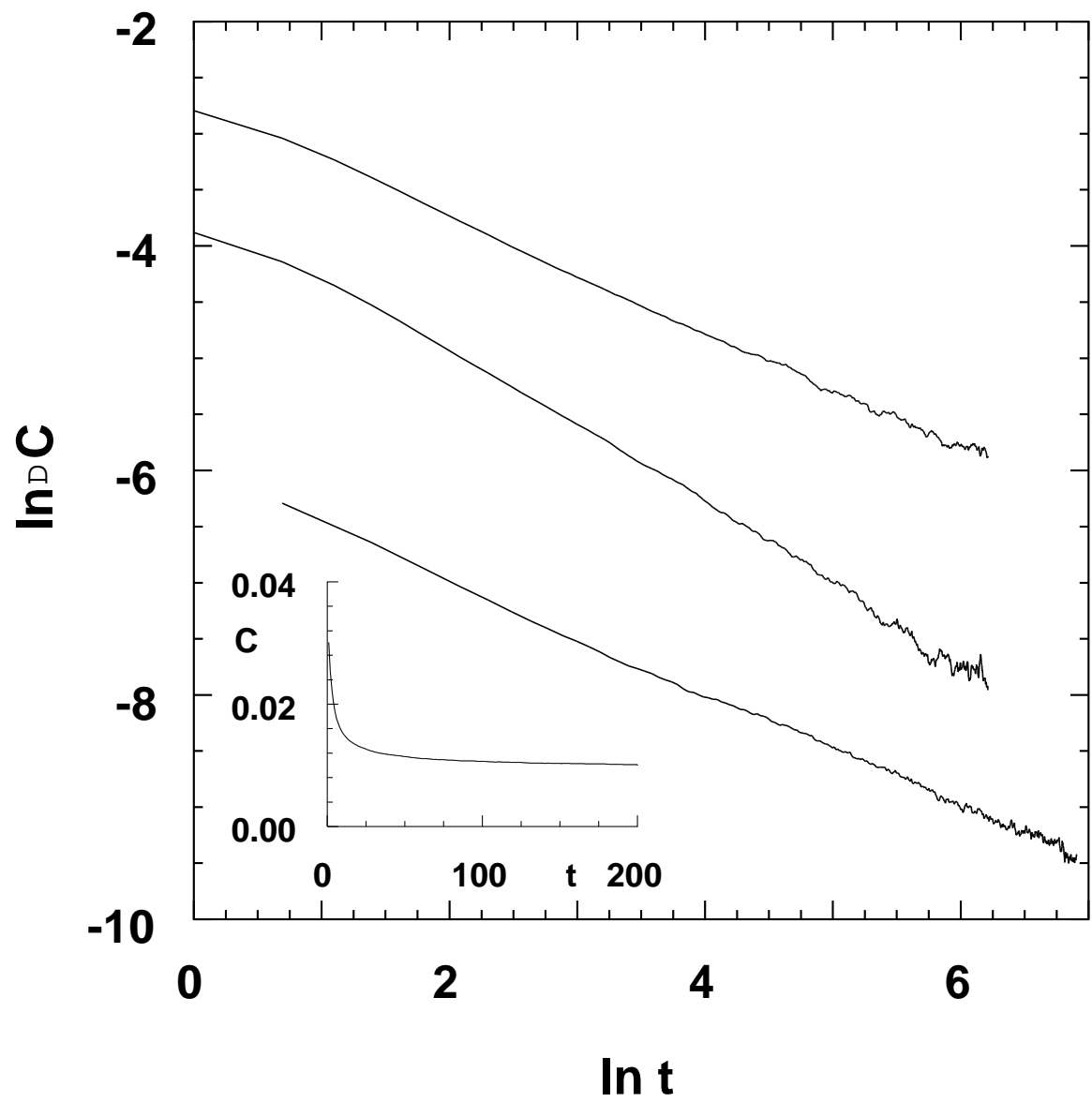
\noindent FIG. 3. Main graph: Excess density-density correlation function versus time
in the DCP for $x=0.3$ and $\lambda = 2.60$ (top), 2.70 (middle), and 2.47 (bottom).
The inset shows C versus t for $x=0.3$ and $\lambda = 2.70$.
\vspace{1em}

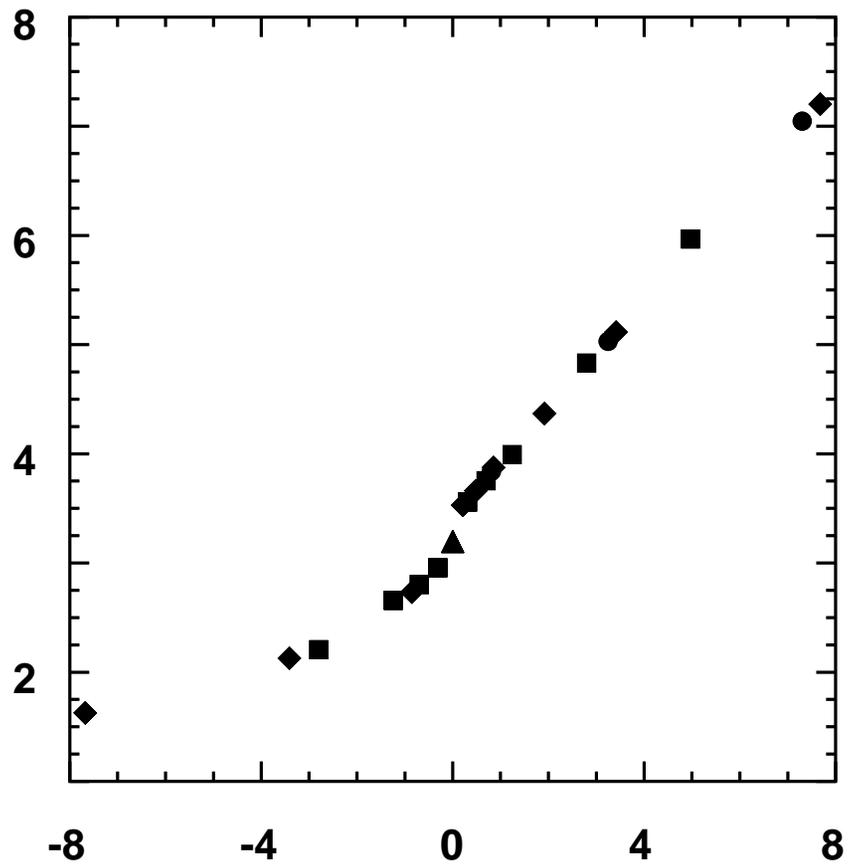
\noindent FIG. 4. Semilogarithmic scaling plot of the lifetime in the {\em pure} CP.
Squares: $|\Delta| = 0.01$; diamonds: $|\Delta| = 0.02$; $\bullet$: $|\Delta| = 0.05$.
\vspace{1em}

\noindent FIG. 5. Survival probability $P_m$ versus time in the critical DCP starting from a
maximally occupied state, for dilution $x=0.1$ and $\lambda = \lambda_c = 1.8464$.
$+$: $L=32$; $\circ$: $L=64$; $\diamond$: $L=128$.
\vspace{1em}

\noindent FIG. 6. Same as Fig. 5, but for $x=0.3$, $\lambda = \lambda_c = 2.47$.
\vspace{1em}

\noindent FIG. 7. Mean number of particles $n_m$ in the critical DCP with $x=0.3$.  Symbols
as in Fig. 5.
\vspace{1em}


\begin{thebibliography}{99}
\bibitem{RDPRV} R. Dickman, in {\em Nonequilibrium
Statistical Mechanics in One Dimension}, V. Privman, Ed. (Cambridge Press, 1996).
\bibitem{TEHARRIS} T. E. Harris, Ann. of Prob. {\bf 2}, 969 (1974).
\bibitem{KINZEL} W. Kinzel, Z. Phys. {\bf B58}, 229 (1985).
\bibitem{NOEST} A. J. Noest, Phys. Rev. Lett. {\bf 57}, 90 (1986).
\bibitem{DCP1} A.G. Moreira and R. Dickman, Phys. Rev. E{\bf 54}, R3090 (1996).
\bibitem{GRASS} P. Grassberger and A. de la Torre, Ann. Phys. (N.Y.) {\bf 122},
 373 (1979).
\bibitem{JANSSEN97} H. K. Janssen, Phys. Rev. E{\bf 55}, 6253 (1997).
\bibitem{betanote} In the CP and other models with a unique absorbing state, $\beta' = \beta$,
but for models with an infinite number of absorbing states $\beta'$ is nonuniversal, as
discussed in Ref. \cite{MENDES}.
\bibitem{MENDES} J.F.F. Mendes, R. Dickman, M. Henkel, and M.C. Marques, J. Phys. 
A{\bf 27}, 3019 (1994).
\bibitem{MUNOZ} M.A. Mu\~{n}oz, G. Grinstein, R. Dickman, and R. Livi,  
Phys. Rev. Lett. {\bf 76}, 451 (1996). 
\bibitem{CHAYES} J. T. Chayes, L. Chayes, D. S. Fisher, and T. Spencer,  
Phys. Rev. Lett. {\bf 57}, 2999 (1986). 
\bibitem{NOEST88} A. J. Noest, Phys. Rev. B {\bf 38}, 2715 (1988).
\bibitem{BRAY} A. J. Bray, Phys. Rev. Lett. {\bf 60}, 720 (1988).
\bibitem{WISEMAN} S. Wiseman and E. Domany, Phys. Rev. E. {\bf 52}, 3469 (1995).
\bibitem{AHARONI} A. Aharoni and A. B. Harris, Phys. Rev. Lett. {\bf 77}, 3700 (1996).

 
\end{thebibliography}
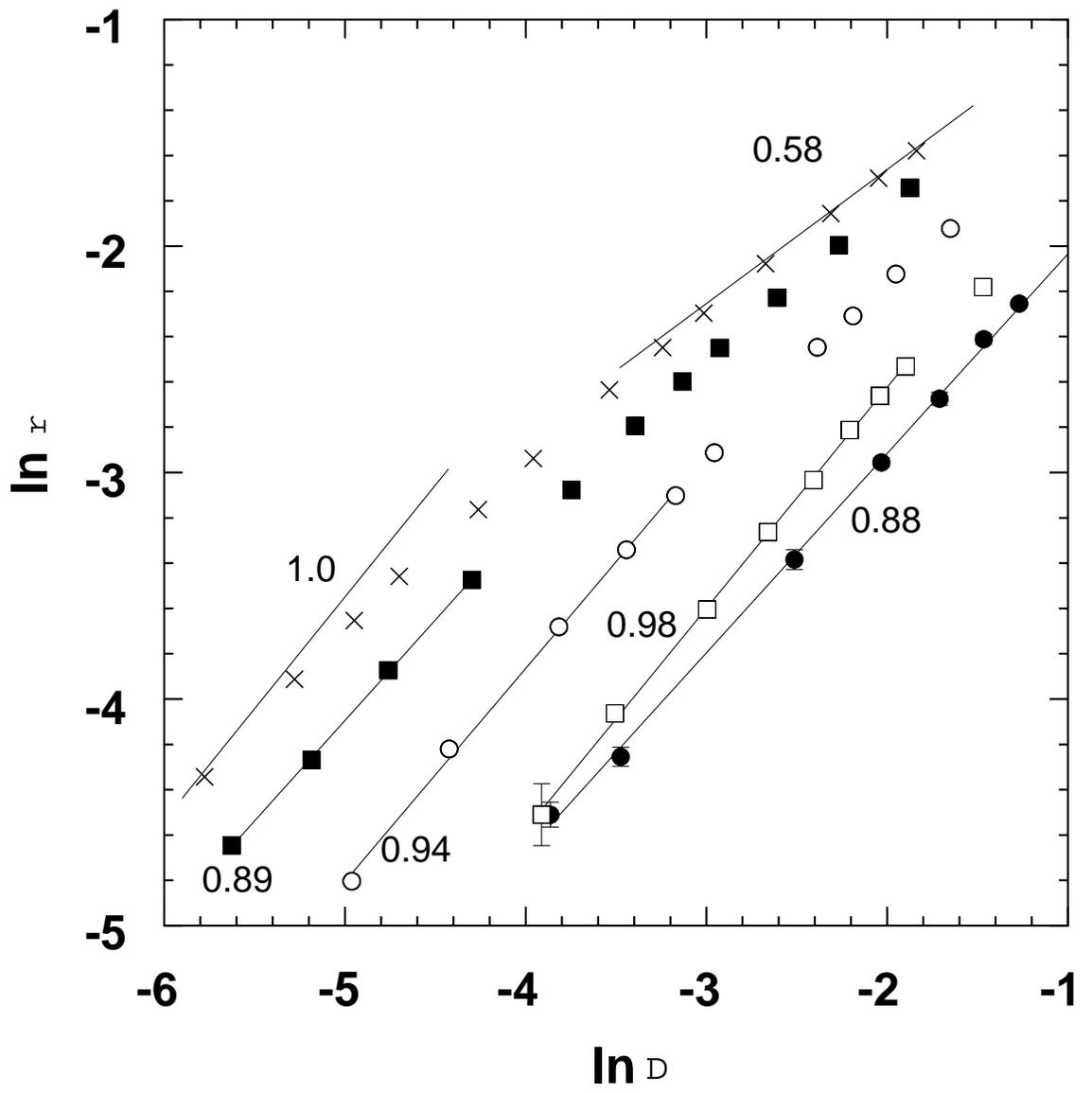
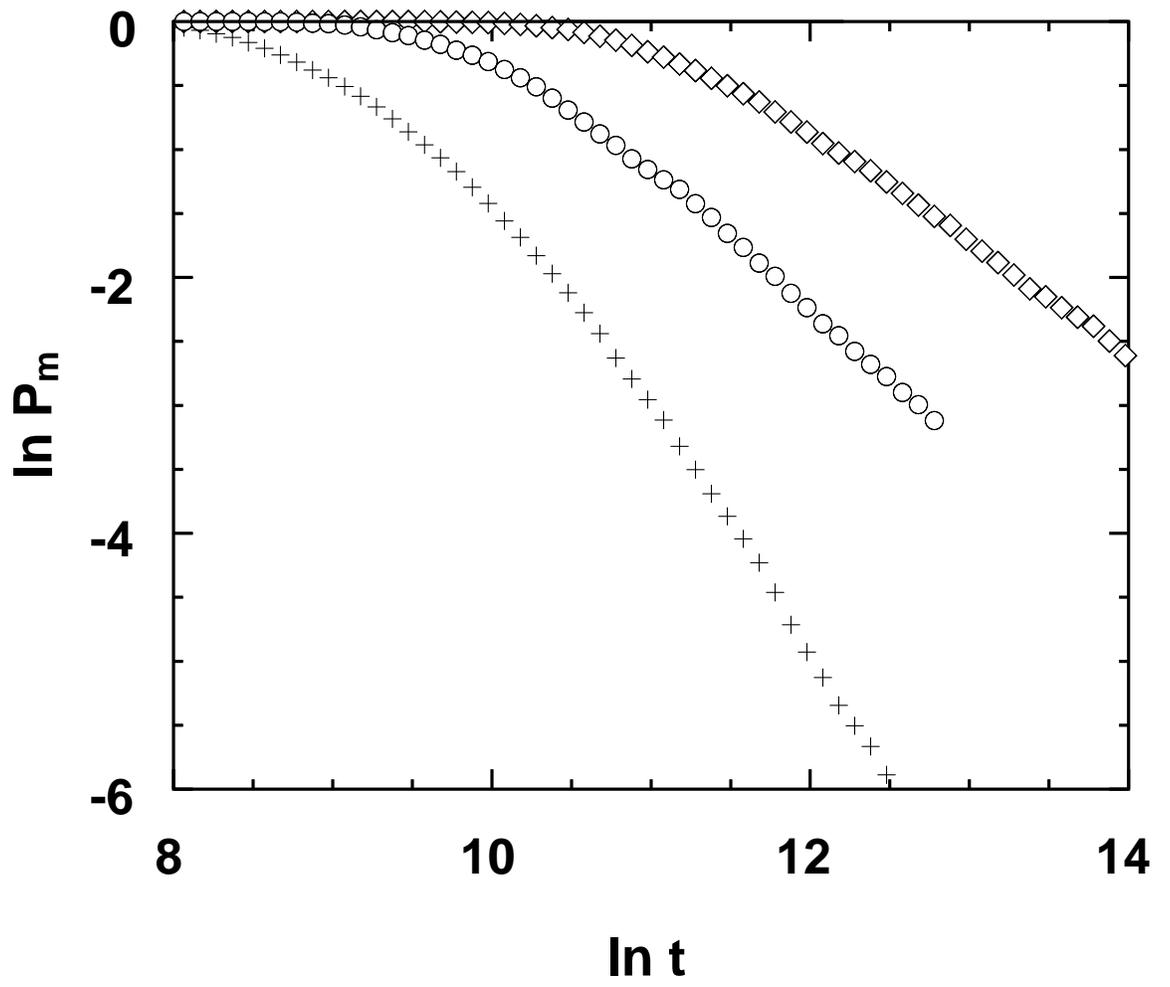
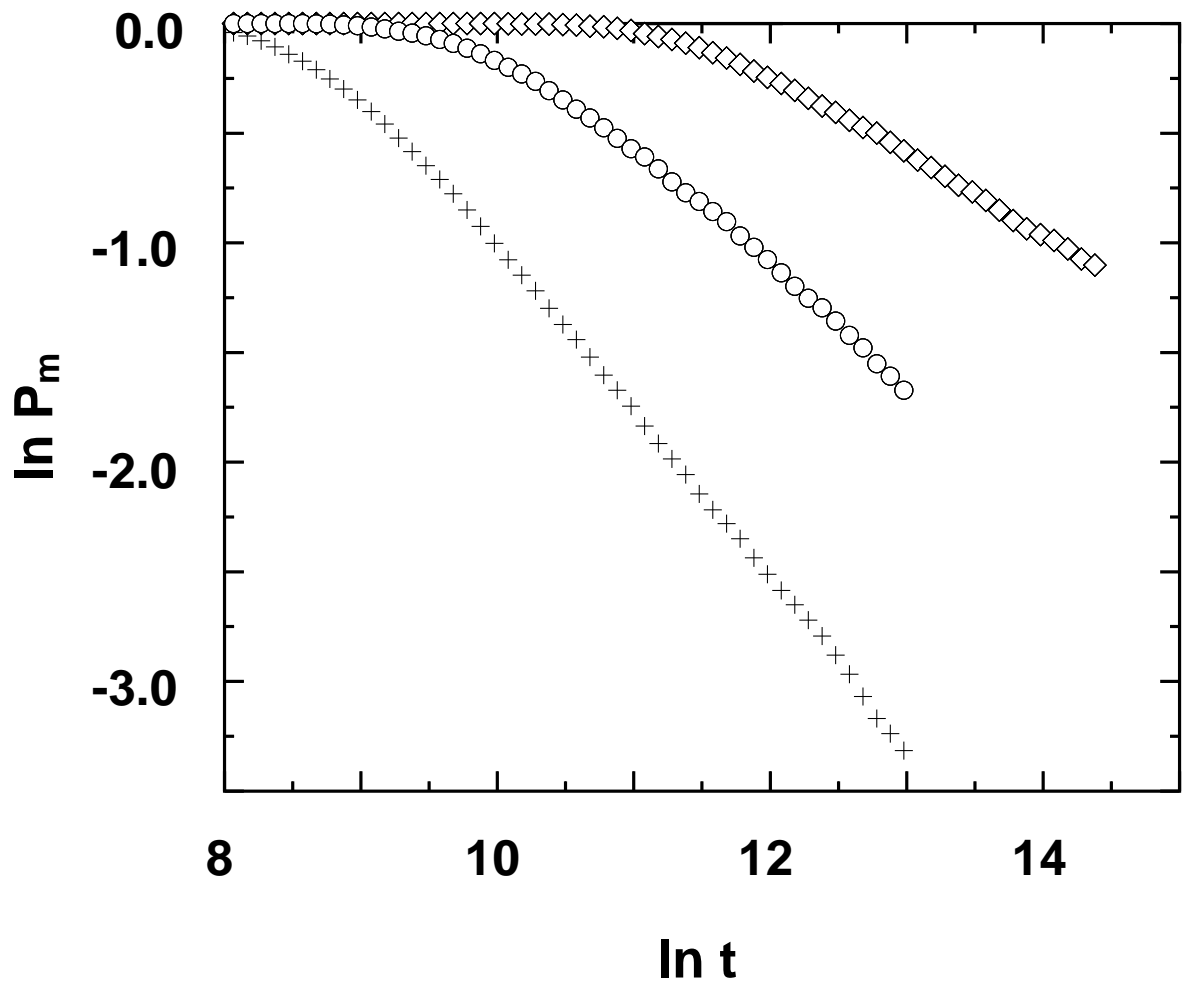
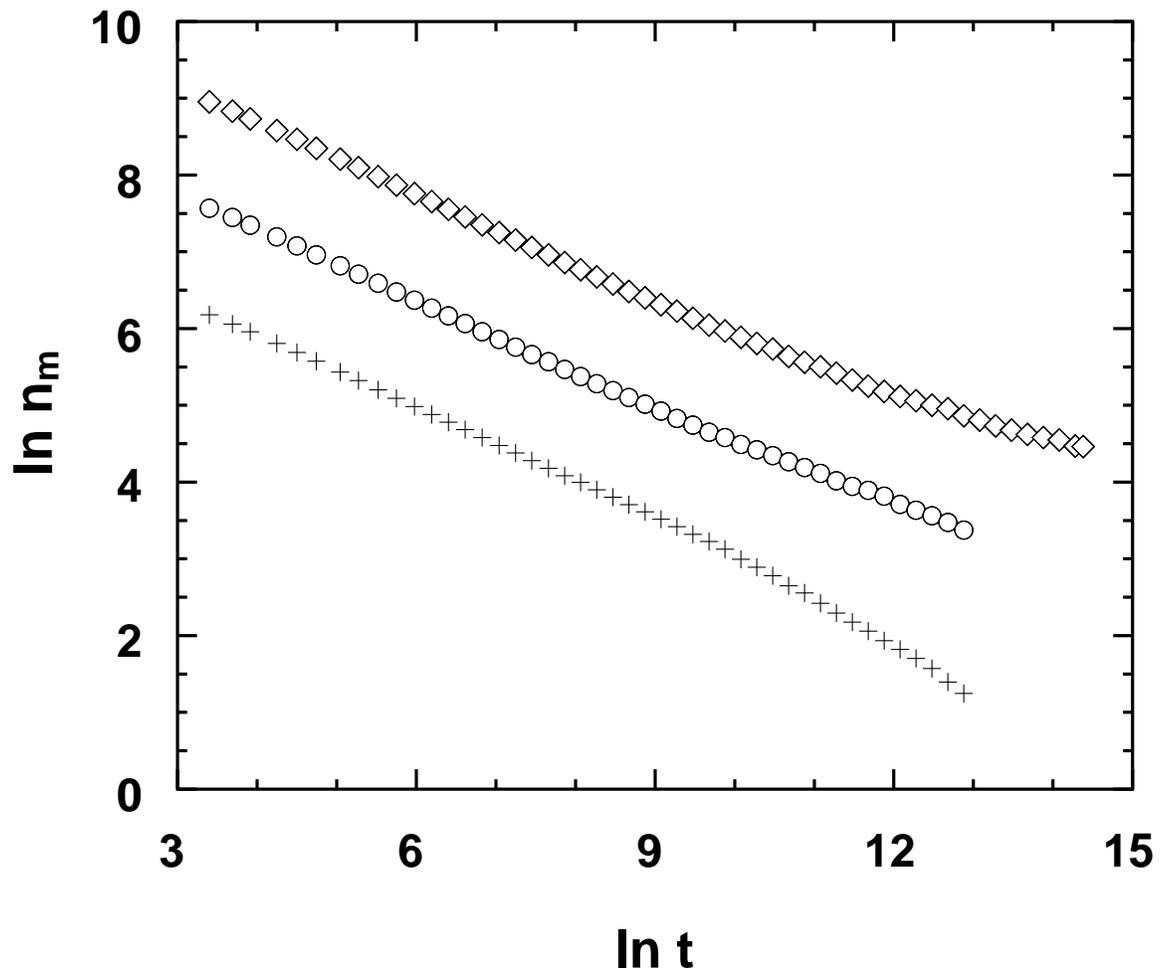
\end{document}